# Degrees of Freedom of the MIMO Interference Channel with Cooperation and Cognition

Chiachi Huang and Syed A. Jafar

Electrical Engineering and Computer Science

University of California Irvine

Irvine, CA, USA

Email: {chiachih, syed}@uci.edu

## Abstract

In this paper, we explore the benefits, in the sense of total (sum rate) degrees of freedom (DOF), of cooperation and cognitive message sharing for a two-user multiple-input-multiple-output (MIMO) Gaussian interference channel with $M_1$, $M_2$ antennas at transmitters and $N_1$, $N_2$ antennas at receivers. For the case of cooperation (including cooperation at transmitters only, at receivers only, and at transmitters as well as receivers), the DOF is $\min\{M_1 + M_2, N_1 + N_2, \max(M_1, N_2)), \max(M_2, N_1)\}$, which is the same as the DOF of the channel without cooperation. For the case of cognitive message sharing, the DOF is $\min\{M_1 + M_2, N_1 + N_2, (1 - 1_{T2})((1 - 1_{R2})\max(M_1, N_2) + 1_{R2}(M_1 + N_2)), (1 - 1_{T1})((1 - 1_{R1})\max(M_2, N_1) + 1_{R1}(M_2 + N_1))\}$ where $1_{Ti} = 1$ (0) when transmitter $i$ is (is not) a cognitive transmitter and $1_{Ri}$ is defined in the same fashion. Our results show that while both techniques may increase the sum rate capacity of the MIMO interference channel, only cognitive message sharing can increase the DOF. We also find that it may be more beneficial for a user to have a cognitive transmitter than to have a cognitive receiver.

## I. Introduction

Multiple-input-multiple-output (MIMO) systems have been proven to be very powerful in point-to-point communication. Following their success in the point-to-point case, MIMO techniques have been widely applied to various multiuser communication scenarios. Since the capacity region for most network communication scenarios has been an open question for many years, capacity approximations are needed to provide an evaluation of the system performance. The number of degrees of freedom (DOF), which is also known as capacity pre-log or multiplexing gain [1], provides a capacity approximation $C_\Sigma(\rho_\Sigma) = \eta \log(\rho_\Sigma) + o(log(\rho_\Sigma))$ where $\eta$ is the number of degrees of freedom, $C_\Sigma(\rho_\Sigma)$ is the sum rate capacity, and $\rho_\Sigma$ is the signal-to-noise ratio (the total transmit power of all nodes divided by the local noise power). The approximation error is within $o(\log(\rho_\Sigma))$ for any $\rho_\Sigma$ and the accuracy of the approximation approaches 100% as $\rho_\Sigma$ increases. The DOF of various multiuser MIMO systems have been found. The two-user multiple access channel (MAC) with $M_1$, $M_2$ antennas at the transmitters and $N$ antenna at the receiver has the DOF of $\min(M_1 + M_2, N)$ [2]. The two-user broadcast channel (BC) with $M$ antennas at the transmitter and $N_1$, $N_2$ antennas at the receivers has the DOF of $\min(M, N_1 + N_2)$ [3]–[5]. The DOF of the two-user MIMO interference channel with $M_1$, $M_2$ antennas at the transmitter and $N_1$, $N_2$ antennas at the receivers, which will be referred to $(M_1, M_2, N_1, N_2)$ interference channel later in this paper, is $\min\{M_1 + M_2, N_1 + N_2, \max(M_1, N_2), \max(M_2, N_1)\}$ [6]. Note that in the MIMO MAC and BC, the distributed processing at either the transmitter side or the receiver side does not cause any loss in the DOF. But in the MIMO interference channel, there may be a significant loss in the DOF due to the distributed processing at both transmitter and receive sides. For example, while a $(1, n, n, 1)$ interference channel has only one degrees of freedom, the point-to-point MIMO system with $1 + n$ antennas at both transmitter and receiver has $1 + n$ degrees of freedom. Many techniques are possible candidates to compensate the loss in DOF caused by the distributed processing nature of the MIMO interference channel. In this paper, we consider two of them: user cooperation and cognitive message sharing. We will explore the benefits, in the sense of DOF, of these two techniques for a two-user MIMO Gaussian interference channel.

### A. Cooperation

The basic idea for cooperation is that several nodes cooperate with each other and act as a large virtual antenna array. Nodes can cooperate to form a transmit antenna array or receive antenna array. Cooperation is made possible by allowing noisy links between distributed transmitters or distributed receivers. A two-user interference channel with single antenna at all nodes is considered by Host-Madsen and Nosratinia in [7], [8]. They show that the number of DOF is equal to one when cooperation takes place at transmitters only, at receivers only, or both at transmitters as well as receivers. However, the DOF for the two-user MIMO interference channel with cooperation remains unknown. One of the goals that we pursue in this paper is to answer this question. We find an upper bound for the DOF of the two-user MIMO Gaussian interference channel with cooperation. The upper bound coincides with the DOF of the channel without cooperation. Thus, we obtain the negative result that cooperation can not increase the DOF of the two-user MIMO Gaussian interference channel, a generalized result from the single antenna case.

### B. Cognitive Message Sharing

Cognitive message sharing refers to genie-aided cooperation in the manner of cognitive radio. In the cognitive radio model, some messages are made available to some nodes (other than the intended nodes) non-causally, noiselessly, and for free [9]. The nodes that get the shared messages are called either cognitive transmitters or cognitive receivers depending on their roles in the channel. Cooperation among users for the interference channel with single antenna at all nodes has been studied in [10]–[13] in the context of cognitive radio channel. The DOF for a $(M, M, M, M)$ interference channel with cognitive message sharing has been studied in [14]. They find that cognitive message sharing can increase the DOF of the channel for some cognitive scenarios. They also find that there is no difference, in the sense of DOF, for a user to have a cognitive transmitter or to have a cognitive receiver. However, the corresponding DOF result and the difference between having a cognitive transmitter and a cognitive receiver for a more general $(M_1, M_2, N_1, N_2)$ interference channel remain unknown. The second goal of this paper is to find the DOF along with the DOF region of a $(M_1, M_2, N_1, N_2)$ interference channel with various cognitive message sharing scenarios. We find that the total number of DOF of a $(M_1, M_2, N_1, N_2)$ interference channel is given by

$$\eta_{1_{T1}, 1_{T2}, 1_{R1}, 1_{R2}} = \min \left\{ \begin{array}{l} M_1 + M_2 \\ N_1 + N_2 \\ (1 - 1_{T2}) \left\{ (1 - 1_{R2}) \max(M_1, N_2) + 1_{R2}(M_1 + N_2) \right\} \\ (1 - 1_{T1}) \left\{ (1 - 1_{R1}) \max(M_2, N_1) + 1_{R1}(M_2 + N_1) \right\} \end{array} \right\} \tag{1}$$

where $1_{Ti} = 1$ if transmitter $i$ is a cognitive transmitter and $1_{Ti} = 0$ if transmitter $i$ is not a cognitive transmitter and $1_{Ri}$ is defined in the same fashion. Our results show that in general, it may be more beneficial, in the sense of DOF, for a user to have a cognitive transmitter than to have cognitive receiver.

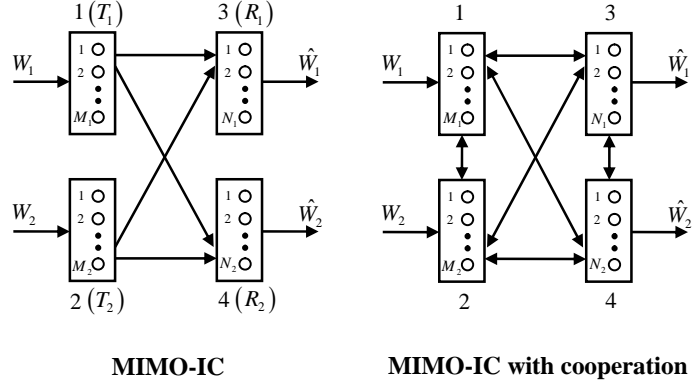

Fig. 1.  Channel models for MIMO interference channels with and without cooperation.

## II. SYSTEM MODEL

A two-user Gaussian MIMO interference channel (MIMO-IC) is defined by

$$\begin{aligned} \mathbf{Y}^{[3]} &= \mathbf{H}^{[31]}\mathbf{X}^{[1]} + \mathbf{H}^{[32]}\mathbf{X}^{[2]} + \mathbf{N}^{[3]} \\ \mathbf{Y}^{[4]} &= \mathbf{H}^{[41]}\mathbf{X}^{[1]} + \mathbf{H}^{[42]}\mathbf{X}^{[2]} + \mathbf{N}^{[4]} \end{aligned} \tag{2}$$

where $\mathbf{Y}^{[3]}$ is the $N_1 \times 1$ output vector at the node 3, $\mathbf{Y}^{[4]}$ is the $N_2 \times 1$ output vector at node 4, $\mathbf{X}^{[1]}$ is the $M_1 \times 1$ input vector at node 1, $\mathbf{X}^{[2]}$ is the $M_2 \times 1$ input vector at node 2, $\mathbf{N}^{[3]}$ is the $N_1 \times 1$ additive white Gaussian noise (AWGN) vector at node 3, $\mathbf{N}^{[4]}$ is the $N_2 \times 1$ additive white Gaussian noise (AWGN) vector at node 4, and $\mathbf{H}^{[ji]}$ is the channel matrix from node $i$ to $j$. All vectors and matrices are real. We assume that all channel matrices are fixed and known to all transmitters and receivers. We also assume that all channel coefficients values are drawn from a continuous distribution. This assumption ensures that all channel matrices are full rank with probability one. Furthermore, the transmitters are subject to an average transmit power $\rho$.

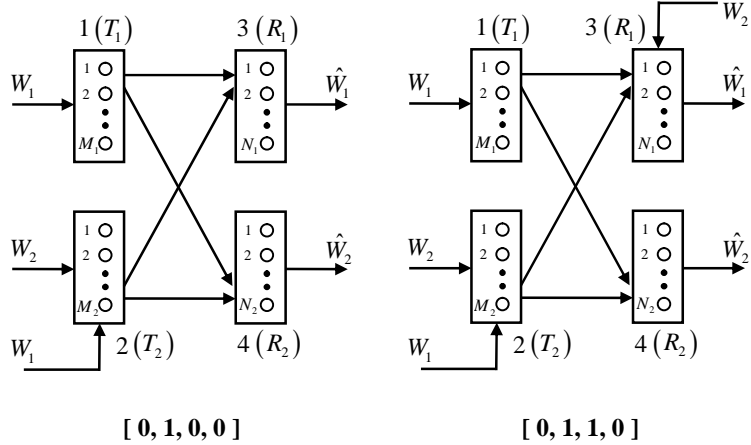

Fig. 2.  Channel models for MIMO interference channels with cognition. Two scenarios are shown in which $[1_{T1}, 1_{T2}, 1_{R1}, 1_{R2}] = [0, 1, 0, 0]$ and $[1_{T1}, 1_{T2}, 1_{R1}, 1_{R2}] = [0, 1, 1, 0]$ separately.

There are two independent messages in the channel: $W_1$ and $W_2$ where $W_i$ is the intended message from node $i$ to node $i+2$, $i = 1, 2$. The message sets are assumed to be functions of $\rho$, and we indicate the size of the message set by $|W_i(\rho)|$. For codewords spanning $n$ channel uses, the rate $R_i(\rho) = \frac{\log |W_i(\rho)|}{n}$ is achievable if the probability of error for $W_i$ can be made arbitrarily small. The capacity region $\mathcal{C}(\rho)$ of the channel is defined as the set of all simultaneously achievable rate tuples $\mathbf{R}(\rho) = (R_1(\rho), R_2(\rho))$. Similar to the definition of the degrees of freedom region in [14], we define the degrees of freedom region $\mathcal{D}$ of the Gaussian MIMO-IC as

$$\mathcal{D} \triangleq \left\{ (d_1, d_2) \in \mathbb{R}_+^2 : \forall (w_1, w_2) \in \mathbb{R}_+^2 \right.$$

$$\left. w_1 d_1 + w_2 d_2 \quad \leq \quad \limsup_{\rho \to \infty} \left( \sup_{\mathbf{R}(\rho) \in \mathcal{C}(\rho)} \frac{w_1 R_1(\rho) + w_2 R_2(\rho)}{\log(\rho)} \right) \right\}$$

The (total) degrees of freedom $\eta$ of the Gaussian MIMO-IC is defined as

$$\eta \triangleq \max_{\mathcal{D}} (d_1 + d_2).$$

We use the following notational conventions. The convex hull of the set $A$ is denoted by $\mathrm{co}(A)$. The function $\max(x, 0)$ is denoted by $(x)^+$. $\mathbb{R}_+^n$ and $\mathbb{Z}_+^n$ represent the sets of n-tuples of non-negative real numbers and integers respectively.

## III. Degrees of Freedom of the MIMO Interference Channel with Cognition

In this section, we find the DOF along with the DOF region of a $(M_1, M_2, N_1, N_2)$ interference channel with various cognitive message sharing scenarios. We use the term "cognitive message sharing" to refer to the message sharing in the manner of cognitive radio. We let $1_{Ti} = 1$ (0) to indicate that transmitter $i$ is (is not) a cognitive transmitter. $1_{Ri}$ is defined in the same fashion. There are total 16 possible combinations of cognitive message sharing scenarios. Figure 2 gives some examples of the possible combinations. Note that in our model, node 1 is transmitter 1, node 2 is transmitter 2, node 3 is receiver 1, and node 4 is receiver 2. A specific cognitive message sharing scenario is labeled by $[1_{T1}, 1_{T2}, 1_{R1}, 1_{R2}]$. We use $\eta_{1_{T1}, 1_{T2}, 1_{R1}, 1_{R2}}$ and $\mathcal{D}_{1_{T1}, 1_{T2}, 1_{R1}, 1_{R2}}$ to denote the DOF and the DOF region of scenario $[1_{T1}, 1_{T2}, 1_{R1}, 1_{R2}]$. We start from an achievable scheme.

*Definition 1:* Define $\mathcal{A}_{1_{T1}, 1_{T2}, 1_{R1}, 1_{R2}}$ to be the set of all $(d_1, d_2) \in \mathbb{Z}_+^2$ satisfying

$$1_{T1}M_1 + M_2 \geq 1_{T1}d_1 + d_2$$
$$M_1 + 1_{T2}M_2 \geq d_1 + 1_{T2}d_2$$
$$N_1 \geq (1 - 1_{R1})(d_2 - (1_{T1}M_1 + M_2 - N_1)^+)^+ + d_1$$
$$N_2 \geq (1 - 1_{R2})(d_1 - (M_1 + 1_{T2}M_2 - N_2)^+)^+ + d_2.$$

The following theorem provides an inner bound for $\mathcal{D}_{1_{T1}, 1_{T2}, 1_{R1}, 1_{R2}}$.

*Theorem 1:*

$$\mathcal{D}_{1_{T1}, 1_{T2}, 1_{R1}, 1_{R2}}^{in} \triangleq \mathrm{co}(\mathcal{A}_{1_{T1}, 1_{T2}, 1_{R1}, 1_{R2}}) \subseteq \mathcal{D}_{1_{T1}, 1_{T2}, 1_{R1}, 1_{R2}}.$$

*Proof:* First, we show that any $(d_1, d_2) \in \mathcal{A}_{1_{T1}, 1_{T2}, 1_{R1}, 1_{R2}}$ is achievable. Instead of providing a proof for general scenario $[1_{T1}, 1_{T2}, 1_{R1}, 1_{R2}]$, we prove the achievability for the scenario $[0, 1, 0, 1]$ to illustrate the key ideas and avoid the tediousness and complexity of dividing cases in the general scenario. Let

$$r_1 = (M_1 + 1_{T2}M_2 - N_2)^+ = (M_1 + M_2 - N_2)^+$$
$$r_2 = (1_{T1}M_1 + M_2 - N_1)^+ = (M_2 - N_1)^+.$$

Choose $\mathbf{v_1^{[31]}}, \ldots, \mathbf{v_{r_1}^{[31]}} \in \mathbb{R}_+^{M_1}$ and $\mathbf{v_1^{[32]}}, \ldots, \mathbf{v_{r_1}^{[32]}} \in \mathbb{R}_+^{M_2}$ such that

$$\left[ \begin{array}{cc} \mathbf{H^{[41]}} & \mathbf{H^{[42]}} \end{array} \right] \left[ \begin{array}{ccc} \mathbf{v_1^{[31]}} & \ldots & \mathbf{v_{r_1}^{[31]}} \\ \mathbf{v_1^{[32]}} & \ldots & \mathbf{v_{r_1}^{[32]}} \end{array} \right] = \left[ \begin{array}{ccc} \mathbf{0} & \ldots & \mathbf{0} \end{array} \right]$$

When $d_1 \leq r_1$, only $\mathbf{v_1^{[31]}}, \ldots, \mathbf{v_{d_1}^{[31]}}$ and $\mathbf{v_1^{[32]}}, \ldots, \mathbf{v_{d_1}^{[32]}}$ are needed. When $d_1 > r_1$, choose the remaining $\mathbf{v_{r_1+1}^{[31]}}, \ldots, \mathbf{v_{d_1}^{[31]}}$ and $\mathbf{v_{r_1+1}^{[32]}}, \ldots, \mathbf{v_{d_1}^{[32]}}$ according to an isotropic distribution so that the set

$$\mathcal{S}_1 = \left\{ \left[ \begin{array}{c} \mathbf{v_1^{[31]}} \\ \mathbf{v_1^{[32]}} \end{array} \right], \ldots, \left[ \begin{array}{c} \mathbf{v_{d_1}^{[31]}} \\ \mathbf{v_{d_1}^{[32]}} \end{array} \right] \right\}$$

is linearly independent with probability one. Choose $\mathbf{v_1^{[42]}}, \ldots, \mathbf{v_{r_2}^{[42]}} \in \mathbb{R}_+^{M_2}$ such that

$$\mathbf{H^{[42]}} \left[ \begin{array}{ccc} \mathbf{v_1^{[42]}} & \ldots & \mathbf{v_{r_2}^{[42]}} \end{array} \right] = \left[ \begin{array}{ccc} \mathbf{0} & \ldots & \mathbf{0} \end{array} \right].$$

When $d_2 \leq r_2$, we only need $\mathbf{v_1^{[42]}}, \ldots, \mathbf{v_{d_2}^{[42]}}$. When $d_2 > r_2$, choose the remaining $\mathbf{v_{r_2+1}^{[42]}}, \ldots, \mathbf{v_{d_2}^{[42]}}$ according to an isotropic distribution so that the set

$$\mathcal{S}_2 = \left\{ \mathbf{v_1^{[42]}}, \ldots, \mathbf{v_{d_2}^{[42]}} \right\}$$

is linearly independent with probability one. Note that since all $\mathbf{v_i^{[31]}}$, $\mathbf{v_j^{[32]}}$, and $\mathbf{v_k^{[42]}}$ are chosen separately and we require (implicitly or explicitly) $d_1 + d_2 \leq M_1 + M_2$, the set

$$\mathcal{S} = \mathcal{S}_1 \bigcup \left\{ \begin{bmatrix} \mathbf{0} \\ \mathbf{v_1^{[42]}} \end{bmatrix}, \ldots, \begin{bmatrix} \mathbf{0} \\ \mathbf{v_{d_2}^{[42]}} \end{bmatrix} \right\}$$

is linearly independent with probability one. After choosing all transmit vectors, let

$$\mathbf{X^{[1]}} = \sum_{i=1}^{d_1} \mathbf{v_i^{[31]}} x_i^{[1]} \tag{3}$$

$$\mathbf{X^{[2]}} = \sum_{i=1}^{d_1} \mathbf{v_i^{[32]}} x_i^{[1]} + \sum_{i=1}^{d_2} \mathbf{v_i^{[42]}} x_i^{[2]} \tag{4}$$

where $x_i^{[j]}$ represents the $i_{th}$ input used to transmit the codeword for message $W_j$.

$$\mathbf{Y^{[3]}} = \underbrace{\sum_{i=1}^{d_1} x_i^{[1]} \left( \mathbf{H^{[31]}} \mathbf{v_i^{[31]}} + \mathbf{H^{[32]}} \mathbf{v_i^{[32]}} \right)}_{\text{range space dimension} = d_1} + \underbrace{\sum_{i=1}^{r_2} x_i^{[2]} \mathbf{H^{[32]}} \mathbf{v_i^{[42]}}}_{=0} + \underbrace{\sum_{i=r_2+1}^{d_2} x_i^{[2]} \mathbf{H^{[32]}} \mathbf{v_i^{[42]}}}_{\text{range space dimension} = (d_2 - r_2)^+} + \mathbf{N^{[3]}}$$

In order to provide enough dimensions to separate the intended signals and the interference, the achievable scheme requires that

$$N_1 \geq d_1 + (d_2 - (M_2 - N_1)^+)^+.$$

Note that among all $N_1$ dimensions at node 3, there are $d_1$ dimensions for the intended signals and $(d_2 - r_2)^+$ dimensions for the interference. By discarding the dimensions that contain the interference, there are $d_1$ degrees of freedom for $W_1$. We want to point out that the dimensions of the intersection of the signal space and the interference space is zero with probability one.

Since node 4 is a cognitive receiver, it can subtract all the signals that caries $W_1$. So we only need $N_2 \geq d_2$ to obtain $d_2$ degrees of freedom for $W_2$. Thus, $(d_1, d_2)$ is achievable. By time sharing, $\text{co}(\mathcal{A}_{0,1,0,1})$ is achievable. ∎

We need the following lemma for the converse.

*Lemma 2:* For any $(d_1, d_2) \in \mathcal{D}_{1_{T_1}, 1_{T_2}, 1_{R_1}, 1_{R_2}}$, the following statements are true.

$$L_1: \quad d_1 + d_2 \leq \min(M_1 + M_2, N_1 + N_2)$$

$$L_2: \quad d_1 \leq N_1$$

$$L_3: \quad d_2 \leq N_2$$

$$L_4: \quad \text{If } 1_{T2} = 0, \text{ then } d_1 \leq M_1$$

$$L_5: \quad \text{If } 1_{T1} = 0, \text{ then } d_2 \leq M_2$$

$$L_6: \quad \text{If } 1_{T2} 1_{R2} = 0, \text{ then } d_1 + d_2 \leq \max(M_1, N_2)$$

$$L_7: \quad \text{If } 1_{T1} 1_{R1} = 0, \text{ then } d_1 + d_2 \leq \max(M_2, N_1)$$

*Proof:* $L_1$ is trivial. $L_2$ and $L_4$ ($L_3$ and $L_5$) are obtained by letting $W_2$ ($W_1$) be a dummy message that is known priori for all nodes. We refer $L_6$ and $L_7$ to Theorem 1 and Corollary 1 in [6]. Note that in the proof of Theorem 1 in [6], the message is provided by a genie to a receiver. But the result is actually stronger in the sense that even the message is given to both the transmitter and the receiver of the same user, all arguments in the proof still hold. ∎

*Corollary 3:* Define $\mathcal{D}^{out}_{1_{T_1},1_{T_2},1_{R_1},1_{R_2}}$ as the set of all $(d_1, d_2) \in \mathbb{R}^2_+$ that satisfy $L_1$ to $L_7$ in Lemma 2. Then

$$\mathcal{D}_{1_{T_1},1_{T_2},1_{R_1},1_{R_2}} \subseteq \mathcal{D}^{out}_{1_{T_1},1_{T_2},1_{R_1},1_{R_2}}.$$

*Theorem 4:*

$$\mathcal{D}^{in}_{1_{T_1},1_{T_2},1_{R_1},1_{R_2}} = \mathcal{D}_{1_{T_1},1_{T_2},1_{R_1},1_{R_2}} = \mathcal{D}^{out}_{1_{T_1},1_{T_2},1_{R_1},1_{R_2}}.$$

*Proof:* Again, we provide the proof for scenario [0, 1, 0, 1] to illustrate the key ideas. Using the fact that $d_1 \leq N_1$ and $d_2 \leq N_2$ ensure that $d_1 + d_2 \leq N_1 + N_2$, we can remove the constraint $d_1 + d_2 \leq N_1 + N_2$ in $\mathcal{D}^{out}_{0,1,0,1}$. Reorganizing the constraints in $\mathcal{D}^{out}_{0,1,0,1}$, we have the following

$$\mathcal{D}^{out}_{0,1,0,1} = \left\{ (d_1, d_2) \in \mathbb{R}^2_+ : \begin{array}{l} d_1 \leq N_1 \\ d_2 \leq \min(M_2, N_2) \\ d_1 + d_2 \leq \min(M_1 + M_2, \max(M_2, N_1)) \end{array} \right\}$$

Using Lemma 5 bellow, the constraint $N_1 \geq d_1 + (d_2 - (M_2 - N_1)^+)^+$ in $\mathcal{A}_{0,1,0,1}$ is equivalent to $d_1 \leq N_1$ and $d_1 + d_2 \leq \max(M_2, N_1)$. Reorganizing the constraints in $\mathcal{A}_{0,1,0,1}$, we find that $\mathcal{A}_{0,1,0,1} = \mathcal{D}^{out}_{0,1,0,1} \cap \mathbb{Z}^2_+$. Observing the constraints in $\mathcal{A}_{0,1,0,1}$ (or $\mathcal{D}^{out}_{0,1,0,1}$), we can find that all intersections of the boundaries take place at points where x-coordinate and y-coordinate are both nonnegative integers. Therefore, we have

$$\mathcal{D}^{in}_{0,1,0,1} = \mathrm{co}(\mathcal{A}_{0,1,0,1}) = \mathcal{D}^{out}_{0,1,0,1}.$$

Following the similar procedure, one can prove that the theorem holds for all scenarios. ∎

*Lemma 5:* For all $a, b, c, d \in \mathbb{Z}^2_+$,

$$\left\{ (a, b) : a + (b - (c - d)^+)^+ \leq d \right\} = \left\{ (a, b) : \begin{array}{l} a \leq d \\ a + b \leq \max(c, d) \end{array} \right\}$$

*Theorem 6:* $\eta_{1_{T_1},1_{T_2},1_{R_1},1_{R_2}}$ is given by (1).

*Proof:* The theorem is proved by solving the linear programming $\max_{\mathcal{D}_{1_{T_1},1_{T_2},1_{R_1},1_{R_2}}}(d_1 + d_2)$ for each case. ∎

*Corollary 7:*

$$\mathcal{D}_{0,0,0,1} \subseteq \mathcal{D}_{0,1,0,0} = \mathcal{D}_{0,1,0,1} \subseteq \mathcal{D}_{0,1,1,0} \subseteq \mathcal{D}_{1,1,0,0} \tag{5}$$

$$\eta_{0,0,0,1} \leq \eta_{0,1,0,0} = \eta_{0,1,0,1} \leq \eta_{0,1,1,0} \leq \eta_{1,1,0,0} \tag{6}$$

Some interesting observation can be drawn for the corollary. First, it may be more powerful, in the sense of DOF, for a user to have a cognitive transmitter than to have a cognitive receiver. Second, for a specific user, after having a cognitive transmitter, having a cognitive receiver does not increase the DOF.

## IV. Degrees of Freedom of the MIMO Interference Channel with Cooperation

In this section, we find the DOF of a $(M_1, M_2, N_1, N_2)$ interference channel with cooperation among users.

*A. System Model*

Cooperation among users is made possible by allowing noisy links between users. In order to provide these noisy links, the system model for the $(M_1, M_2, N_1, N_2)$ MIMO-IC defined in (2) is generalized to

$$\mathbf{Y}^{[i]}(n) = \sum_{j=1}^{4} \mathbf{H}^{[ij]}\mathbf{X}^{[j]}(n) + \mathbf{N}^{[i]}(n) \tag{7}$$

where $n$ is the index for time slot and the definitions of $\mathbf{X}^{[i]}$, $\mathbf{Y}^{[i]}$, $\mathbf{H}^{[ij]}$, and $\mathbf{N}^{[i]}$ are similar to those in Section II. Note that in our new model, all nodes are allowed to transmit and receive in full duplex mode. But there are still only two messages (as before) - $W_1$ from node 1 to node 3 and $W_2$ from node 2 to node 4. All nodes are subject to an average transmit power $\rho$. We define $\mathbf{X}^{[i]^n}$ as

$$\mathbf{X}^{[i]^n} \triangleq \left[\ \mathbf{X}^{[1]}(\mathbf{1})\ \ \dots\ \ \mathbf{X}^{[i]}(\mathbf{n})\ \right]^t.$$

Similar definitions apply to $\mathbf{Y}^{[i]^n}$ and $\mathbf{Z}^{[i]^n}$. The encoding and decoding functions are

$$\mathbf{X}^{[i]}(n) = f_{1,n}\left(W_i, \mathbf{Y}^{[i]^{(n-1)}}\right)$$
$$\mathbf{X}^{[i+2]}(n) = f_{i+2,n}\left(\mathbf{Y}^{[i+2]^{(n-1)}}\right)$$
$$\hat{W}_{i+2} = g_{i+2}\left(\mathbf{Y}^{[i+2]^N}\right)$$

where $N$ is the codewords length and for $i = 1, 2$.

*B. Main Results*

In order to find the upper bound of the DOF of the $(M_1, M_2, N_1, N_2)$ interference channel with cooperation among users, we define the auxiliary random variables $\mathbf{U}^{[1]}(n)$, $\mathbf{U}^{[2]}(n)$, $\mathbf{U}^{[3]}(n)$, $\mathbf{U}^{[4]}(n)$ as

$$\mathbf{U}^{[i]}(n) = \mathbf{H}^{[i1]}\mathbf{X}^{[1]}(n) + \mathbf{N}^{[i]}(n), \quad i = 1, 2, 3, 4$$

The following lemma is needed to prove our main theorem.

*Lemma 8:* These statements are true.

$$
\begin{aligned}
L_1: & \quad \mathbf{X}^{[1]^n} & \leftarrow W_1, W_2, \mathbf{U}^{[1]^{n-1}}, \mathbf{U}^{[2]^{n-1}}, \mathbf{U}^{[3]^{n-1}}, \mathbf{U}^{[4]^{n-1}} \\
L_2: & \quad \mathbf{X}^{[2]^n}, \mathbf{X}^{[3]^n}, \mathbf{X}^{[4]^n} & \leftarrow W_2, \mathbf{U}^{[1]^{n-1}}, \mathbf{U}^{[2]^{n-1}}, \mathbf{U}^{[3]^{n-1}}, \mathbf{U}^{[4]^{n-1}} \\
L_3: & \quad \mathbf{Y}^{[1]^n}, \mathbf{Y}^{[2]^n}, \mathbf{Y}^{[3]^n}, \mathbf{Y}^{[4]^n} & \leftarrow W_2, \mathbf{U}^{[1]^n}, \mathbf{U}^{[2]^n}, \mathbf{U}^{[3]^n}, \mathbf{U}^{[4]^n}
\end{aligned}
$$

where $A \leftarrow B$ denotes that $A$ can be completely determined by $B$.

Next, we provide a genie-based upper bound for the DOF of the $(M_1, M_2, N_1, N_2)$ MIMO-IC with cooperation where $N_2 \geq M_1$. Before providing the theorem, we like to mention that the proof is an extension from the single antenna setting in [8] and [15] to the MIMO setting. While the extension is straightforward for the most part, we include it for the sake of completeness.

*Theorem 9:* When $N_2 \geq M_1$, the DOF of the $(M_1, M_2, N_1, N_2)$ MIMO-IC with cooperation satisfies

$$\eta \leq N_2.$$

*Proof:* Suppose that a genie provides node 3 with side information containing $W_2$, $\mathbf{U}^{[1]^n}$, $\mathbf{U}^{[2]^n}$, $\mathbf{U}^{[3]^n}$, and $\mathbf{U}^{[4]^n}$. According to lemma 8, node 3 can construct $\mathbf{Y}^{[i]^n}$, $i = 1, 2, 3, 4$ using the side information. Using Fano's inequality we have the following

$$
\begin{aligned}
NR_1(\rho) &\leq I\left(W_1; W_2, \mathbf{Y}^{[3]N}, \mathbf{U}^{[1]N}, \mathbf{U}^{[2]N}, \mathbf{U}^{[3]N}, \mathbf{U}^{[4]N}\right) + N\epsilon_N \tag{8} \\
&= I\left(W_1; W_2, \mathbf{U}^{[1]N}, \mathbf{U}^{[2]N}, \mathbf{U}^{[3]N}, \mathbf{U}^{[4]N}\right) + N\epsilon_N \tag{9} \\
&= I\left(W_1; \mathbf{U}^{[1]N}, \mathbf{U}^{[2]N}, \mathbf{U}^{[3]N}, \mathbf{U}^{[4]N} \mid W_2\right) + N\epsilon_N \tag{10} \\
&= \underbrace{H\left(\mathbf{U}^{[1]N}, \mathbf{U}^{[2]N}, \mathbf{U}^{[3]N}, \mathbf{U}^{[4]N} \mid W_2\right)}_{T_1} - \underbrace{H\left(\mathbf{U}^{[1]N}, \mathbf{U}^{[2]N}, \mathbf{U}^{[3]N}, \mathbf{U}^{[4]N} \mid W_1, W_2\right)}_{T_2} + N\epsilon_N \tag{11}
\end{aligned}
$$

where $T_1$ can be expressed and bounded above as follow

$$
\begin{aligned}
T_1 &= \sum_{n=1}^{N} H\left(\mathbf{U}^{[1]}(n), \mathbf{U}^{[2]}(n), \mathbf{U}^{[3]}(n), \mathbf{U}^{[4]}(n) \mid W_2, \mathbf{U}^{[1]n-1}, \mathbf{U}^{[2]n-1}, \mathbf{U}^{[3]n-1}, \mathbf{U}^{[4]n-1}\right) \tag{12} \\
&= \sum_{n=1}^{N} H\left(\mathbf{U}^{[4]}(n) \mid W_2, \mathbf{U}^{[1]n-1}, \mathbf{U}^{[2]n-1}, \mathbf{U}^{[3]n-1}, \mathbf{U}^{[4]n-1}\right) \\
&\quad + \sum_{n=1}^{N} H\left(\mathbf{U}^{[1]}(n), \mathbf{U}^{[2]}(n), \mathbf{U}^{[3]}(n) \mid W_2, \mathbf{U}^{[1]n-1}, \mathbf{U}^{[2]n-1}, \mathbf{U}^{[3]n-1}, \mathbf{U}^{[4]n-1}, \mathbf{U}^{[4]}(n)\right) \tag{13} \\
&\stackrel{(a)}{=} \sum_{n=1}^{N} H\left(\mathbf{U}^{[4]}(n) + \sum_{i=2}^{4} \mathbf{H}^{[4i]}\mathbf{X}^{[i]}(n) \mid W_2, \mathbf{U}^{[1]n-1}, \mathbf{U}^{[2]n-1}, \mathbf{U}^{[3]n-1}, \mathbf{U}^{[4]n-1}\right) \\
&\quad + \sum_{n=1}^{N} H\left(\mathbf{U}^{[1]}(n), \mathbf{U}^{[2]}(n), \mathbf{U}^{[3]}(n) \mid W_2, \mathbf{U}^{[1]n-1}, \mathbf{U}^{[2]n-1}, \mathbf{U}^{[3]n-1}, \mathbf{U}^{[4]n-1}, \mathbf{U}^{[4]}(n)\right) \tag{14} \\
&\stackrel{(b)}{=} \sum_{n=1}^{N} H\left(\mathbf{Y}^{[4]}(n) \mid W_2, \mathbf{U}^{[1]n-1}, \mathbf{U}^{[2]n-1}, \mathbf{U}^{[3]n-1}, \mathbf{U}^{[4]n-1}, \mathbf{Y}^{[4]n-1}\right) \\
&\quad + \sum_{n=1}^{N} H\left(\mathbf{U}^{[1]}(n), \mathbf{U}^{[2]}(n), \mathbf{U}^{[3]}(n) \mid W_2, \mathbf{U}^{[1]n-1}, \mathbf{U}^{[2]n-1}, \mathbf{U}^{[3]n-1}, \mathbf{U}^{[4]n-1}, \mathbf{U}^{[4]}(n)\right) \tag{15} \\
&\stackrel{(c)}{\leq} \sum_{n=1}^{N} H\left(\mathbf{Y}^{[4]}(n) \mid W_2, \mathbf{Y}^{[4]n-1}\right) + \sum_{n=1}^{N} H\left(\mathbf{U}^{[1]}(n), \mathbf{U}^{[2]}(n), \mathbf{U}^{[3]}(n) \mid \mathbf{U}^{[4]}(n)\right) \tag{16} \\
&\stackrel{(d)}{=} H\left(\mathbf{Y}^{[4]N} \mid W_2,\right) + \sum_{n=1}^{N} H\left(\mathbf{U}^{[1]}(n), \mathbf{U}^{[2]}(n), \mathbf{U}^{[3]}(n) \mid \mathbf{U}^{[4]}(n)\right) \tag{17} \\
&\stackrel{(e)}{\leq} H\left(\mathbf{Y}^{[4]N} \mid W_2,\right) + \sum_{n=1}^{N} \sum_{i=1}^{3} H\left(\mathbf{U}^{[i]}(n) \mid \mathbf{U}^{[4]}(n)\right). \tag{18}
\end{aligned}
$$

Equality $(a)$ is obtained by $L_2$ in lemma 8. Equality $(b)$ is obtained by $L_3$ in lemma 8. Inequality $(c)$ and $(e)$ use the fact that conditioning reduces entropy. The second term in (18) can be bounded above by the following method.

We choose $i = 1$ as an example.

$$
\begin{aligned}
H\left(\mathbf{U}^{[1]}(n) \mid \mathbf{U}^{[4]}(n)\right) &= H\left(\mathbf{H}^{[11]}\mathbf{X}^{[1]}(n) + \mathbf{N}^{[1]}(n) \mid \mathbf{H}^{[41]}\mathbf{X}^{[1]}(n) + \mathbf{N}^{[4]}(n)\right) \quad (19) \\
&\overset{(a)}{\leq} \sum_{j=1}^{M_1} H\left(\mathbf{H}_j^{[11]}\mathbf{X}^{[1]}(n) + N_j^{[1]}(n) \mid \mathbf{H}^{[41]}\mathbf{X}^{[1]}(n) + \mathbf{N}^{[4]}(n)\right) \quad (20) \\
&\overset{(b)}{\leq} \sum_{j=1}^{M_1} H\left(\mathbf{H}_j^{[11]}\mathbf{X}^{[1]}(n) + N_j^{[1]}(n) \mid \mathbf{H}_j^{[41]}\mathbf{X}^{[1]}(n) + N_j^{[4]}(n)\right) \quad (21) \\
&\overset{(c)}{\leq} \sum_{j=1}^{M_1} \left(\log\left(1 + \frac{\|\mathbf{H}_j^{[11]}\|^2 \rho}{1 + \|\mathbf{H}_j^{[41]}\|^2 \rho}\right) + \log(2\pi e)\right) \quad (22)
\end{aligned}
$$

where $\mathbf{H}_j^{[i1]}$ and $N_j^{[i]}$ denote the channel and noise associated with the $j_{th}$ antenna at node $i$. Inequality $(a)$ and $(b)$ use the fact that conditioning reduces entropy. We refer inequality $(c)$ to lemma 1 in [7]. We would like to mention that inequality $(d)$ holds only when $\mathbf{H}^{[41]}$ is full rank and $N_2 \geq M_1$ which have been assumed. Next, $T_2$ can be simplified as follow

$$
\begin{aligned}
T_2 &= \sum_{n=1}^{N} H\left(\mathbf{U}^{[1]}(n), \mathbf{U}^{[2]}(n), \mathbf{U}^{[3]}(n), \mathbf{U}^{[4]}(n) \mid W_1, W_2, \mathbf{U}^{[1]^{n-1}}, \mathbf{U}^{[1]^{n-1}}, \mathbf{U}^{[1]^{n-1}}, \mathbf{U}^{[1]^{n-1}}\right) \quad (23) \\
&\overset{(a)}{=} \sum_{n=1}^{N} H\left(\mathbf{U}^{[1]}(n), \mathbf{U}^{[2]}(n), \mathbf{U}^{[3]}(n), \mathbf{U}^{[4]}(n) \mid W_1, W_2, \mathbf{U}^{[1]^{n-1}}, \mathbf{U}^{[1]^{n-1}}, \mathbf{U}^{[1]^{n-1}}, \mathbf{U}^{[1]^{n-1}}, \mathbf{X}^{[1]}(n)\right) \quad (24) \\
&= \sum_{n=1}^{N} H\left(\mathbf{N}^{[1]}(n), \mathbf{N}^{[2]}(n), \mathbf{N}^{[3]}(n), \mathbf{N}^{[4]}(n)\right) \quad (25) \\
&= \sum_{n=1}^{N} \sum_{i=1}^{4} H\left(\mathbf{N}^{[i]}(n)\right) \quad (26) \\
&= N(M_1 + M_2 + N_1 + N_2)\log(2\pi e). \quad (27)
\end{aligned}
$$

Equality $(a)$ is obtained by $L_1$ in lemma 8. The simplification of $T_2$ can be explained as the following. Since $W_1$ and $W_2$ are the only messages in the system, after knowing $W_1$ and $W_2$, the uncertainty in $\mathbf{U}^{[1]^N}$, $\mathbf{U}^{[2]^N}$, $\mathbf{U}^{[3]^N}$, and $\mathbf{U}^{[4]^N}$ is only the uncertain of Gaussian noise. The entropy of Gaussian noise does not increase with $\rho$. Combining (11), (18), (22), and (27), we have

$$
R_1(\rho) \leq \frac{1}{N} H\left(\mathbf{Y}^{[4]^N} \mid W_2,\right) + o(\log(\rho)).
$$

Using Fano's inequality, $R_2(\rho)$ can be bounded above as follow

$$
\begin{aligned}
R_2(\rho) &\leq \frac{1}{N} I\left(W_2; \mathbf{Y}^{[4]^N}\right) + \epsilon_N \quad (28) \\
&= \frac{1}{N} H\left(\mathbf{Y}^{[4]^N}\right) - \frac{1}{N} H\left(\mathbf{Y}^{[4]^N} \mid W_2\right) + \epsilon_N \quad (29)
\end{aligned}
$$

Adding $R_1(\rho)$ and $R_2(\rho)$ together, we get

$$
\begin{aligned}
R_1(\rho) + R_2(\rho) &\leq \frac{1}{N} H\left(\mathbf{Y}^{[4]^N}\right) + o(\log(\rho)) \quad (30) \\
&\leq N_2 \log(\rho) + o(\log(\rho)) \quad (31)
\end{aligned}
$$

where the last equality can be obtained from the property of Gaussian random variable. Thus, we prove that when $N_2 \geq M_1$, the DOF of the system is smaller than equal to $N_2$. ∎

*Corollary 10:* The DOF of the $(M_1, M_2, N_1, N_2)$ MIMO-IC with cooperation satisfies

$$\eta \leq \min \{\max(M_1, N_2), \max(M_2, N_1)\}.$$

*Proof:* If $N_2 \geq M_1$, Theorem 9 can be applied directly to obtain $\eta \leq N_2$. If $M_1 > N_2$, let us add more antennas to node 4 so that node 4 has $M_1$ antennas too. Adding antennas does not hurt, so the converse argument remains. We then apply Theorem 9 to the new MIMO-IC to obtain $\eta \leq M_1$. Thus, we have $\eta \leq \max(M_1, N_2)$ for all possible cases. $\eta \leq \max(M_2, N_1)$ can be obtained by switching indices 1 to 2 and 2 to 1. ∎

Corollary 10 along with Theorem 2 in [6], which gives the DOF of the MIMO-IC without cooperation, lead to the following corollary.

*Corollary 11:* The DOF of the $(M_1, M_2, N_1, N_2)$ MIMO-IC with cooperation satisfies

$$\eta = \min \{M_1 + M_2, N_1 + N_2, \max(M_1, N_2), \max(M_2, N_1)\}.$$

Our result shows that cooperation can't increase the DOF of the MIMO-IC. This result can be thought of as a generalization of the single antenna case in [7].

## V. Conclusions

We investigate the degrees of freedom of the Gaussian MIMO interference channel with cooperation and cognition. We find the general forms of the DOF and the DOF region of the Gaussian MIMO-IC with all possible cognitive message sharing scenarios. Our results show that it may be more powerful for a user to have a cognitive transmitter than to have a cognitive receiver. We also find a negative result that user cooperation does not increase the number of DOF of the Gaussian MIMO-IC. Directions for future work include the generalization from two user case to more user case and the exploration of benefits of having feedback in the setting.